# Plasmonic Heterodyne Spectrometry for Resolving the Spectral Signatures of Ammonia over a 1-5 THz Frequency Range

Yen-Ju Lin[1], Semih Cakmakyapan[1], Ning Wang[1], Daniel Lee[2], Mitchell Spearrin[2], and Mona Jarrahi[1]

[1]*Electrical and Computer Engineering, University of California – Los Angeles, Los Angeles, CA, 90095, USA*

[2]*Mechanical and Aerospace Engineering Department, University of California – Los Angeles, Los Angeles, CA, 90095, USA*

We present a heterodyne terahertz spectrometry platform based on plasmonic photomixing, which enables the resolution of narrow spectral signatures of gases over a broad terahertz frequency range. This plasmonic heterodyne spectrometer replaces the terahertz mixer and local oscillator of conventional heterodyne spectrometers with a plasmonic photomixer and a heterodyning optical pump beam, respectively. The heterodyning optical pump beam is formed by two continuous-wave, wavelength-tunable lasers with a broadly tunable terahertz beat frequency. This broadly tunable terahertz beat frequency enables spectrometry over a broad bandwidth, which is not restricted by the bandwidth limitations of conventional terahertz mixers and local oscillators. We use this plasmonic heterodyne spectrometry platform to resolve the spectral signatures of ammonia over a 1-5 THz frequency range.

**THE MANUSCRIPT**

Heterodyne terahertz spectrometry is an attractive modality for gas sensing, because it can provide high spectral resolution for resolving narrow gas spectral lines [1-6]. It involves background radiation from a terahertz or blackbody source interacting with the gas under test. The radiation received by the heterodyne spectrometer carries the spectral signatures of the gas and is mixed with a terahertz local oscillator signal to downconvert the targeted terahertz spectral signature to an intermediate frequency (IF) signal in the radio frequency (RF) range. The downconverted spectrum is then resolved by backend IF electronics. Schottky diode, superconductor-insulator-superconductor (SIS), and hot electron bolometer (HEB) mixers are used for frequency-downconversion in conventional heterodyne terahertz spectrometers [6-12]. While conventional heterodyne terahertz spectrometers offer high spectral resolution and high sensitivity levels at cryogenic temperatures, their room temperature sensitivity and operation bandwidth are restricted by the sensitivity limitations of room-temperature terahertz mixers and frequency tunability constraints of terahertz local oscillators, respectively.

To address these limitations, we recently introduced a heterodyne terahertz spectrometry scheme based on plasmonic photomixing [13-20]. By replacing the terahertz mixer and local oscillator of conventional heterodyne spectrometers with a plasmonic photomixer and a heterodyning optical pump beam, respectively, we demonstrated a heterodyne terahertz detector with quantum-level sensitivities at room temperature and operation bandwidths exceeding 5 THz [21]. In this work, we use this plasmonic photomixer to resolve the spectral signatures of ammonia. Ammonia gas sensing is of interest for agriculture, combustion exhaust treatment, clinical breath analysis, and industrial process monitoring [22-26]. We chose ammonia for our first spectrometry measurements because its rotational spectra in the terahertz band provide comparable intensity to the strongest infrared vibrational bands. In addition, ammonia has relatively well-isolated lines in the 1-5 THz domain, allowing us to demonstrate the very broad operation bandwidth of the plasmonic heterodyne spectrometer, which cannot be offered by conventional heterodyne spectrometers.

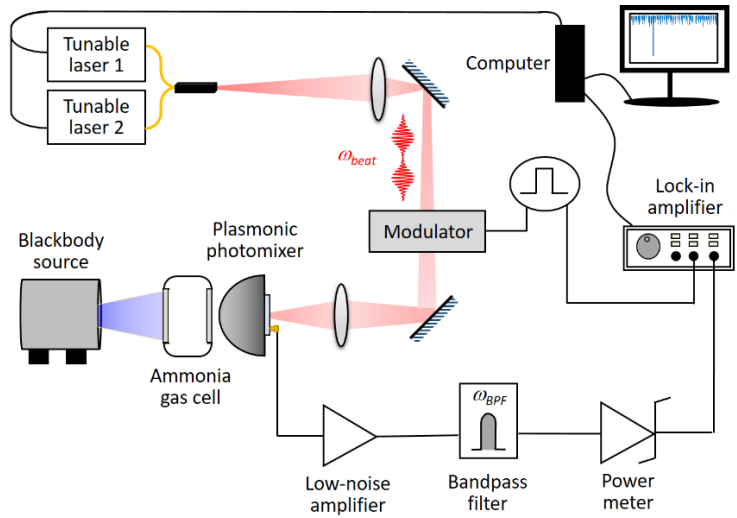

Fig. 1. Schematic diagram of the terahertz spectrometry setup.

A schematic diagram of the experimental setup is shown in Fig. 1. Two continuous-wave, wavelength-tunable lasers with center wavelengths of 780 nm and 785 nm (TOPTICA #DLC-DL-PRO-780 and TOPTICA #LD-0785-0080-DFB-1) are combined to provide the heterodyning optical pump beam with a tunable beat frequency in the 0.1-5 THz range. The heterodyning optical beam is modulated by an acousto-optic modulator (Gooch & Housego AOMO 3080-125) with a 50% duty cycle and a 100 kHz rate and then focused onto the active area of the plasmonic photomixer mounted on a silicon lens [21]. The IF output of the plasmonic photomixer at ~1 GHz is amplified by a low-noise amplifier (Mini-Circuits ZRL-1150) and filtered by a bandpass filter (Mini-Circuits ZVBP-909) with a bandwidth of $BW_{IF}$ = 15 MHz and a center frequency of ~1 GHz. Subsequently, the IF signal is detected by a power meter (Mini-Circuits ZX47-60LN) using a lock-in amplifier with the 100 kHz modulation reference signal, a 2 s integration time, and a 4 kHz bandwidth.

A 1.3-cm-long, 1-inch-diameter, room-temperature ammonia gas cell with two 1.6-mm-thick high-density polyethylene windows (Wavelength References, Inc.) is placed in front of the plasmonic photomixer such that the background blackbody radiation incident on the silicon lens passes through the gas cell. Therefore, the resolved spectra contain spectral dips at the ammonia absorption frequencies. A calibrated blackbody source (IR-563 from Boston Electronics) is placed on the other side of the gas cell to provide the background blackbody radiation. The temperature of the blackbody source is set to 500°C (773 K) for all of the reported spectra in this manuscript. However, it should be noted that even in the absence of the external blackbody source, all of the demonstrated ammonia spectral lines in this manuscript are still observable due to the presence of ambient blackbody radiation at ~300 K.

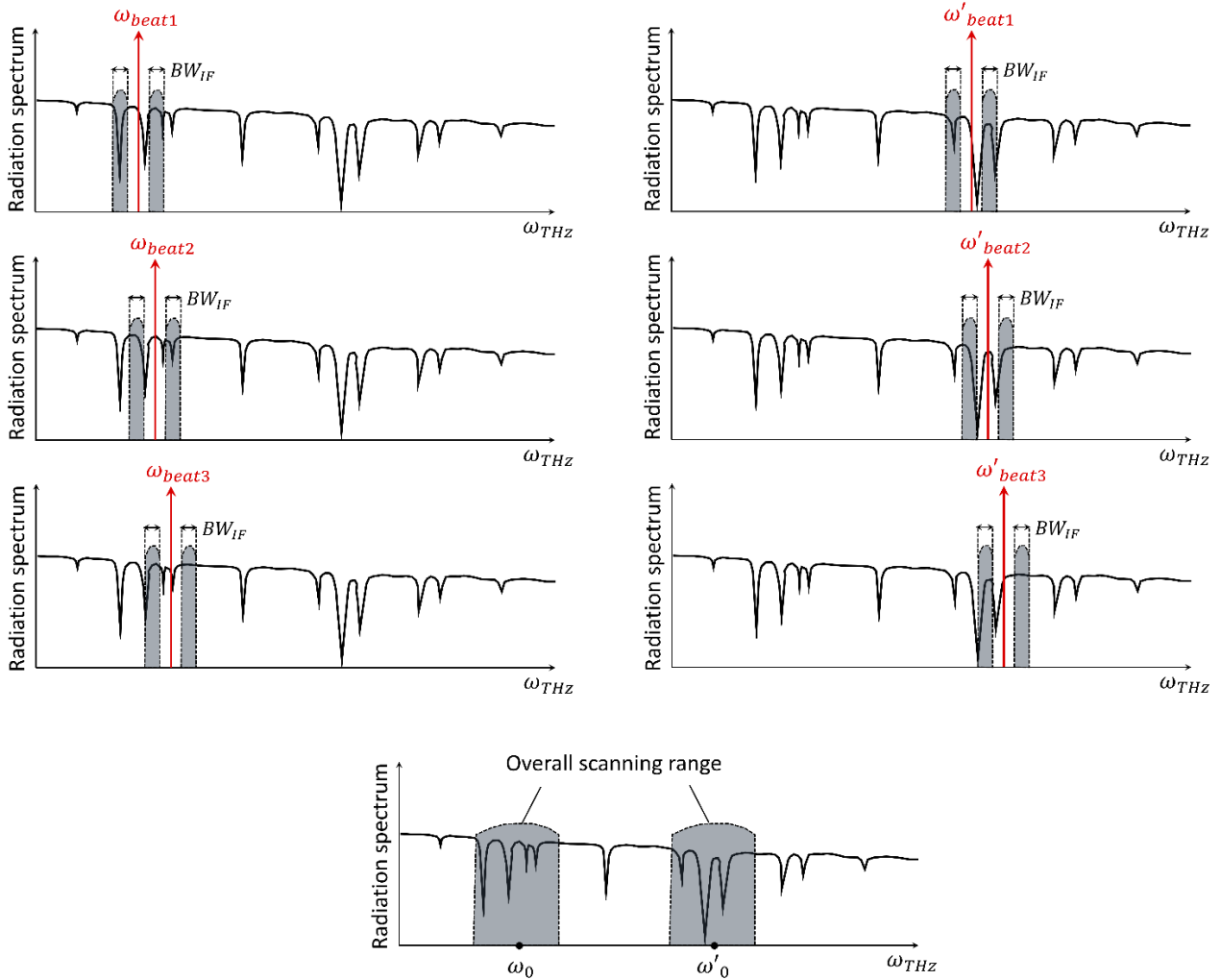

Fig. 2. Broadband heterodyne spectrometry over specific frequency ranges around $\omega_0$ and $\omega'_0$ with high scanning efficiency.

Frequency scanning is performed by varying the center wavelength of one of the tunable lasers (TOPTICA #LD-0785-0080-DFB-1) using a computer program, thereby varying the optical beat frequency, $\omega_{beat}$. The detected IF power at each frequency step, which carries the received spectral information at $\omega_{beat} \pm \omega_{BPF}$ over the bandwidth of the bandpass filter, is recorded by the computer. During all of the spectrometry measurements, the wavelengths of the two laser beams forming the heterodyning optical pump beam are monitored in real time by an optical spectrum analyzer.

One of the advantages of the plasmonic heterodyne terahertz spectrometer is that the scanning frequency range can be limited to specific frequency ranges around the targeted molecular spectral lines. This advantage allows the resolution of molecular spectral signatures over a broad terahertz frequency range with a high scanning efficiency. To achieve this goal, the locations of the spectral lines of the targeted molecule(s) are determined first. Next, as shown in Fig. 2, the pump laser beat frequency is set to each spectral line's center frequency (e.g., $\omega_{beat1}$ and $\omega'_{beat1}$) one after another, and the spectral data over an instantaneous bandwidth equal to the bandwidth of the backend IF electronics, $BW_{IF}$, are resolved. If the bandwidth of the backend IF electronics is not sufficient to cover the entire targeted frequency range around the targeted center frequencies, the pump laser beat frequency is tuned around each spectral line's center frequency (e.g., $\omega_{beat1}$, $\omega_{beat2}$, $\omega_{beat3}$ and $\omega'_{beat1}$, $\omega'_{beat2}$, $\omega'_{beat3}$) one after another with a frequency step smaller than the bandwidth of the backend IF electronics. While the ultimate spectral resolution limit is set by the pump laser linewidth, the resolved spectral resolution is equal to the frequency resolution of the backend IF electronics. By combining the resolved spectra at each optical pump beat frequency, the spectral information is extracted at the desired center frequencies (e.g., $\omega_0$ and $\omega'_0$) over a spectral range that can be significantly broader than the bandwidth of the backend IF electronics. By using this approach, a high scanning efficiency is maintained independent of the locations and linewidths of the molecular spectral lines since the frequency scanning range and step can both be decreased/increased when observing narrow/broad linewidths.

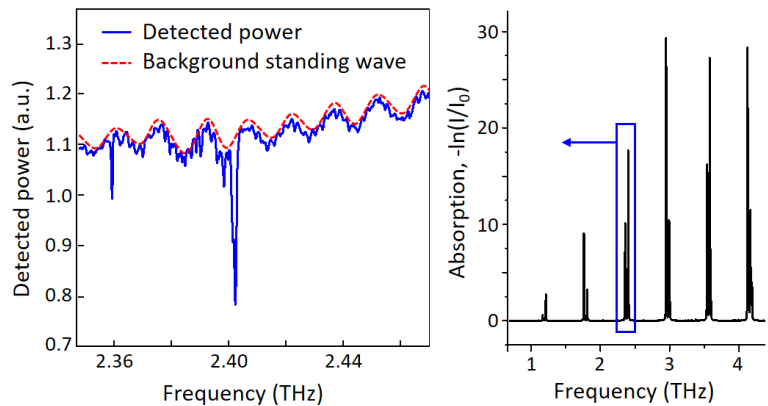

Fig. 3. The detected power spectrum over a 2.35-2.47 THz range. The absorbance spectrum of ammonia is shown on the right [27].

Figure 3 shows the detected power spectrum for a frequency scanning range of 2.35-2.47 THz, which reveals the expected ammonia absorption dips at approximately 2.359 THz and 2.402 THz [27]. The sinusoidal-like background in the spectrum, shown by the red dashed line, is associated with standing waves formed because of the reflections from the gas cell walls. To eliminate the contribution of these standing waves, we have developed a post-processing algorithm that uses a fitting function to extract the background standing waveform from the measured power spectrum. A high-order polynomial fitting function with a least squares error is used to extract the background standing waveform (i.e., the red dashed line shown in Fig. 3). Then, the extracted background standing waveform is subtracted from the measured power spectrum to eliminate the sinusoidal-like background.

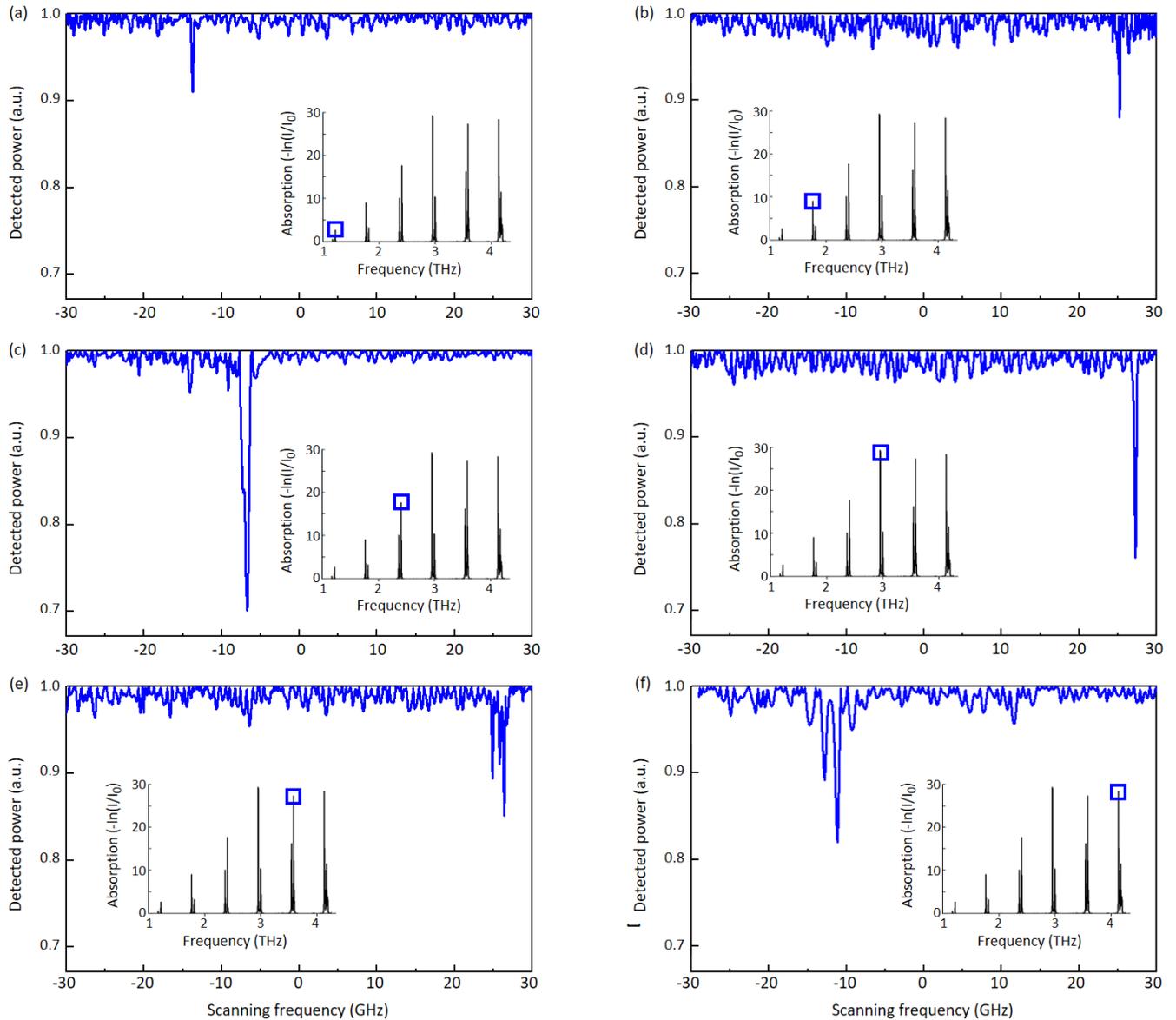

Fig. 4. The resolved power spectra around the ammonia absorption lines at a) 1.215 THz, b) 1.764 THz, c) 2.401 THz, d) 2.950 THz, e) 3.577 THz, and f) 4.125 THz, over a 60 GHz frequency range.

Figure 4 shows the resolved power spectra around the ammonia absorption lines at 1.215 THz, 1.764 THz, 2.401 THz, 2.950 THz, 3.577 THz, and 4.125 THz after post-processing. For each measurement, the optical pump beat frequency is set near these absorption lines, and the spectral information is resolved over a bandwidth of 60 GHz. The scan time for each 60 GHz band is approximately one hour. The 60 GHz frequency scanning range around each ammonia absorption line is marked with a blue box in the inset absorbance spectra. As shown in Fig. 4, all of the targeted ammonia spectral lines in the 1-5 THz frequency range are resolved by only tuning the beat frequency of the optical pump beam. Such a broad frequency scanning range cannot be offered by conventional heterodyne terahertz spectrometers due to the bandwidth limitations of Schottky diode, SIS, and HEB mixers and terahertz local oscillators [28-30]. In analyzing the resolved spectra, the very low amount of ammonia molecules inside the 1.3-cm-long, 1-inch-diameter gas cell should be considered, which leads to relatively low signal-to-noise ratio (SNR) levels. In addition, gas cell calibration in the spectrometry setup would enable the development of more accurate post-processing algorithms that would significantly reduce the background noise of the resolved spectra [31]. Another important factor that would significantly improve the SNR and frequency accuracy of the demonstrated heterodyne spectrometry system is stabilization of the lasers that provide the terahertz beat frequency [32, 33]. To evaluate the impact of laser fluctuations on the stability of the heterodyne spectrometer, the Allan variance of the normalized output power as a function of integration time is calculated for a fixed optical pump beat frequency. Figure 5 shows the calculated Allan variance as a function of integration time for an optical pump beat frequency of 2 THz. For short integration times, the Allan variance is inversely proportional to the integration time and lock-in amplifier bandwidth of 4 kHz, following the radiometric equation. This is the regime in which the noise is dominated by white noise. However, for longer integration times, the drift in the laser output and the 1/f noise dominate, and the Allan variance deviates from the radiometric equation. The increasing Allan variance values for integration times longer than 3 s imply that laser drift is the ultimate stability limit of the spectrometer.

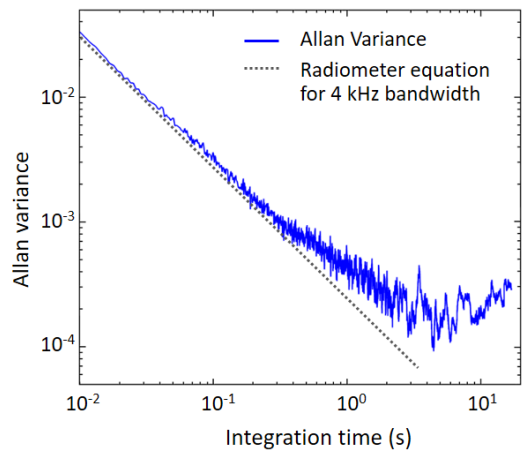

Fig. 5. Allan variance of the normalized output power as a function of integration time for an optical pump beat frequency of 2 THz.

The scanning efficiency of the plasmonic heterodyne terahertz spectrometer can be significantly increased by utilizing backend IF electronics with instantaneous bandwidths larger than 15 MHz and using larger optical beat frequency scanning steps. Different types of backend IF electronics developed for conventional heterodyne spectrometers can be combined with the presented plasmonic photomixer to extract spectral data. These backend IF electronics provide different frequency resolutions and instantaneous bandwidths for resolving different molecular spectral lines [34, 35]. The scanning efficiency can also be increased by using an optical comb laser as the optical pump source to provide multiple optical beat frequencies simultaneously [36-39]. However, since the power of an optical comb is distributed among many comb lines, a small fraction of the total pump power is used for resolving the spectral information at a desired terahertz frequency near the closest optical beat frequency, lowering the IF current corresponding to the desired terahertz frequency. Moreover, although the optical comb lines that produce beat frequencies far from the desired terahertz frequency do not contribute to the IF current corresponding to the desired terahertz frequency, they increase the device noise current [40]. Therefore, the presented spectrometer can offer significantly higher scanning efficiencies by using an optical comb pump at the expense of lower spectrometry sensitivities.

## ACKNOWLEDGMENTS

This work was supported by the Office of Naval Research (contract # N00014-14-1-0573) and National Science Foundation (contract #1305931). Dr. Semih Cakmakyapan was supported by the Department of Energy (grant # DE-SC0016925).